\documentclass{sig-alternate-05-2015}
\setcopyright{acmcopyright}
\pdfoutput=1

\begin{document}
\title{IntelliAd: Understanding In-APP Ad Costs From Users' Perspective}
\numberofauthors{5}
\author{
%
%
\alignauthor
Cuiyun Gao\\
	   \affaddr{Dept. of Computer Sci. \& Erg.}\\
	   \affaddr{The Chinese University of Hong Kong, China}\\
       \email{cygao@cse.cuhk.edu.hk}
\alignauthor
Hui Xu\\ \affaddr{Dept. of Computer Sci. \& Erg.}\\
	  \affaddr{The Chinese University of Hong Kong, China}\\
       \email{hxu@cse.cuhk.edu.hk}
\alignauthor Yichuan Man\\
	   \affaddr{School of Electrical Erg.}\\
       \affaddr{Beijing Jiaotong University, China}\\
       \email{14121444@bjtu.edu.cn}
\and  
\alignauthor Yangfan Zhou\\
       \affaddr{School of Computer Science}\\
       \affaddr{Fudan University, China}\\
       \email{zyf@fudan.edu.cn}
\alignauthor Michael R. Lyu\\
       \affaddr{Dept. of Computer Sci. \& Erg.}\\
	   \affaddr{The Chinese University of Hong Kong, China}\\
       \email{lyu@cse.cuhk.edu.hk}
}

\maketitle
\begin{abstract}
Ads are an important revenue source for mobile app development, especially for free apps, whose expense can be compensated by ad revenue. The ad benefits also carry with costs. For example, too many ads can interfere the user experience, leading to less user retention and reduced earnings ultimately. In the paper, we aim at understanding the ad costs from users' perspective. We utilize app reviews, which are widely recognized as expressions of user perceptions, to identify the ad costs concerned by users. Four types of ad costs, \textit{i.e.}, number of ads, memory/CPU overhead, traffic usage, and bettery consumption, have been discovered from user reviews. To verify whether different ad integration schemes generate different ad costs, we first obtain the commonly used ad schemes from 104 popular apps, and then design a framework named $\mathsf{IntelliAd}$ to automatically measure the ad costs of each scheme. To demonstrate whether these costs indeed influence users' reactions, we finally observe the correlations between the measured ad costs and the user perceptions. We discover that the costs related to memory/CPU overhead and battery consumption are more concerned by users, while the traffic usage is less concerned by users in spite of its obvious variations among different schemes in the experiments. Our experimental results provide the developers with suggestions on better incorporating ads into apps and, meanwhile, ensuring the user experience.

\end{abstract}





\section{Introduction}
Mobile advertising experiences a tremendous growth recently and has already become ubiquitous on mobile terminals~\cite{adreport}. Organizations that have successfully monetized mobile apps generate huge profits. For example, for Twitter, mobile ad revenues have accounted for 88\% of the generated sales in 2014. Underlying the ad benefits is the hidden costs, such as battery drainage and traffic consumption. For example, according to~\cite{mohan2013prefetching}, ads consume 23\% of an app's total energy. Therefore, many research efforts have focused on mitigating the costs, such as prefetching ads to reduce the battery drainage~\cite{mohan2013prefetching}, and using predictive profiling of a user's context to enable lower network overhead~\cite{khan2012mitigating}.

However, the previous studies have ignored that user perceptions are the major indicator of the ad serving quality. They play a decisive role in the ad-profiting process, since ad revenue is calculated based on the count of ads displayed (\textit{impressions}) or clicked (\textit{clicks}) by users. Inappropriate embedding ads into apps, such as displaying too many ads, can interfere users' experience and reduce the user retention. The low user retention then generates less ad display and ultimately cuts down the ad revenue. Thus, the analysis on ad-related user concerns (\textit{i.e.}, ad costs) can benefit both the developers and end users. In the paper, we conduct an \textit{empirical study} to understand the ad costs from the perceptive of users.


Generally, mobile ads delivery includes two types, in-app mobile ads, and mobile web ads. Since the in-app ads generate more 12\% ad impressions than web-based ads~\cite{inappad}, we focus our study on understanding users' perceptions of in-app ads. Furthermore, we mainly consider the ad revenue generated by impressions rather than clicks, for the reason that being seen matters more than being clicked for ad earnings~\cite{addisplay}.

To obtain the user perceptions about ads, we utilize user reviews, which capture unique perspectives about the users' reactions of the apps~\cite{khalid2015mobile}. We have identified four types of ad costs from user reviews, \textit{i.e.}, number of ads, memory/CPU overhead, traffic usage, and battery consumption. Different from the existing work~\cite{gui2015truth,khan2012mitigating}, we do not explore whether the ad-embedded apps can bring users more costs. Instead, we suppose that ad rendering is integral to the development of free apps, and examine whether different ad integration schemes behave differently regarding the ad costs. We aim at providing the software developers with the insights on better embedding ads into apps.

Generally, the developers require to determine the ad network and ad format for incorporating ads into their apps. We refer the combination of these two attributes as an ad integration scheme. To verify the impact of different ad integration schemes on the costs, we have conducted an experimental study on 104 popular apps from Google Play. We first retrieve the ads incorporated into each app based on static analysis. For each ad integration scheme (including the ad network and ad format), we then create a basic prototype app with the ads embedded. We design a framework named $\mathsf{IntelliAd}$ to automatically measure and profile the ad costs of each scheme. We have discovered that different ad integration schemes indeed influence the ad costs differently, and the CPU utilization and traffic usage embody the most obvious differences. To demonstrate whether users truly pay attention to these ad costs, we finally observe the correlations between the measured values and the corresponding user perceptions (\textit{i.e.}, user ratings). We demonstrate that user concerns concentrate more on the costs caused by memory/CPU overhead and battery consumption, but less on the cost produced by traffic usage.

To sum up, we intend to answer the following questions.\par
\begin{inparaenum}[\itshape a\upshape)]
\item What aspects of in-app ads can generate unfavorable perceptions among users? (in Section~\ref{sec:cost})\par
\item Does the ad integration scheme impact the ad costs concerned by users? (in Section~\ref{subsec:analysis})\par
\item Can the measured cost values reflect the user reactions? (in Section~\ref{subsec:users}) \par
\end{inparaenum}

Overall, we try to establish an intense relation between the in-app ad integration and the user experience, and endeavor to provide the developers with suggestions on better incorporating ads into apps. The main contributions of this work are summarized as follows:
\begin{compactitem}
\item We first investigate the ad costs from the users' perspective by analyzing massive user reviews.
\item Our analysis can be easily generalized to other apps with the same ads embedded, since we analyze the ad costs for each ad integration scheme, instead of each app. 
\item We demonstrate the correlations between the ad costs measured and the corresponding user perceptions to provide the app developers more comprehensive understanding about the user concerns.
\end{compactitem}

The remainder of the paper is organized as follows. Section~\ref{sec:review} describes the motivation and the background of our work. Section~\ref{sec:framework} outlines the overall picture. Section~\ref{sec:cost} illustrates the analysis on user reviews and summarizes ad-related user concerns. Section~\ref{sec:implementation} explains static analysis on ad incorporation conditions of real apps, followed by the automatic measurement tool $\mathsf{IntelliAd}$ for ad costs in Section~\ref{sec:intelliad}. The experimental results are described in Section~\ref{sec:experiment}. Section~\ref{sec:discussion} discusses possible limitations. Section~\ref{sec:literature} introduces related work. Finally, Section~\ref{sec:conclusion} concludes the paper.

\section{Motivation and Background}\label{sec:review}
Similar to the ad integration behavior, which is prompted by the benefits carried with ads, our work is also motivated by the ad revenue, specifically, how to optimize the ad revenue for the developers. We consider that users play an essential role in the total ad earnings, and below is our explanation.

\begin{figure}[h]
    \centering
    \includegraphics[width=0.38 \textwidth]{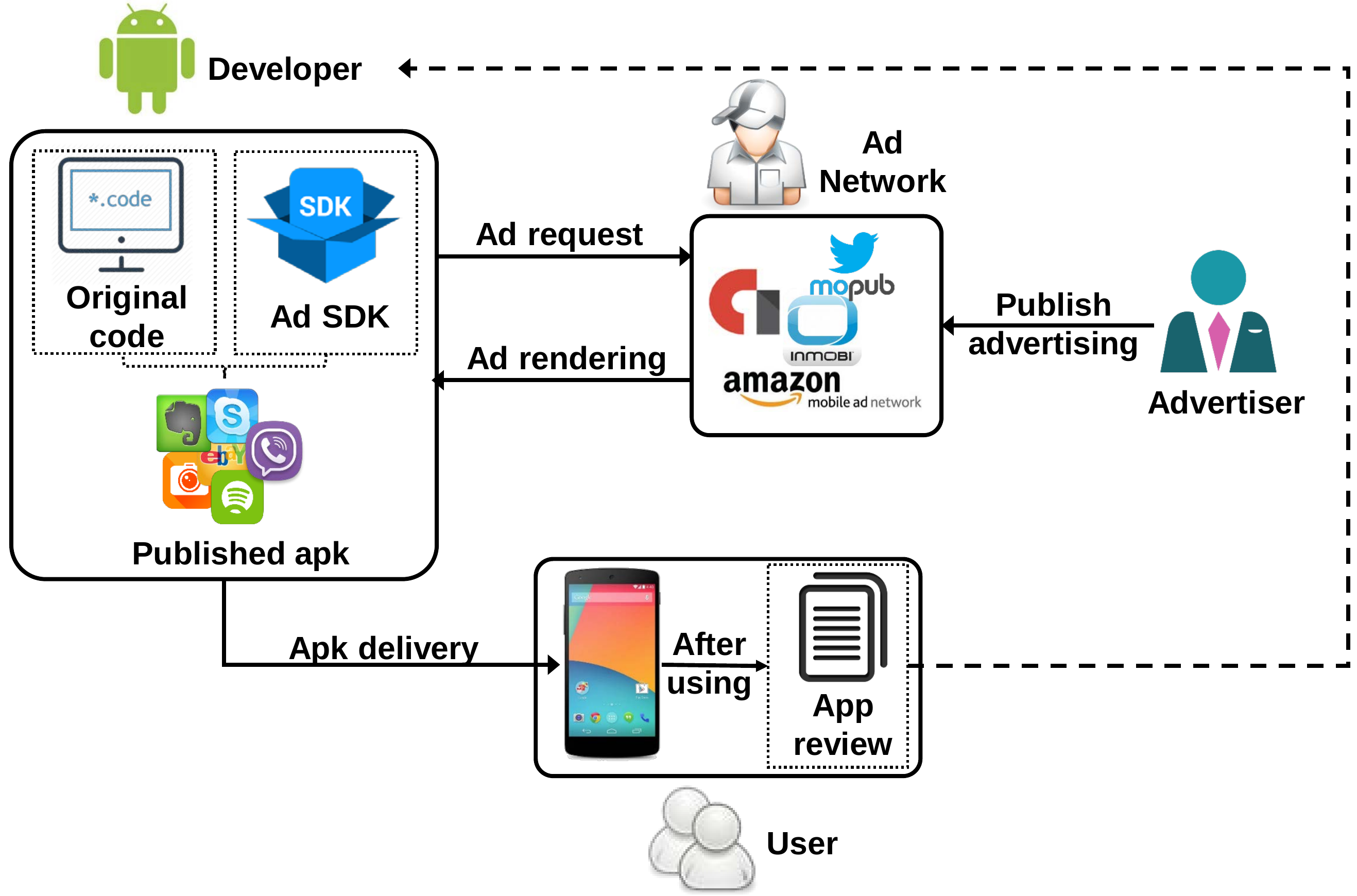}
    \caption{In-App Mobile Advertising Ecosystem.}
\label{fig:adflow}
\end{figure}

For in-app ads, the advertising ecosystem comprises four major ingredients, \textit{i.e.}, app developers, advertisers, ad networks, and, one essential but easily ignored component, end users. The relations among them are illustrated in Figure~\ref{fig:adflow}.

Initially, app software developers are inspired by compensating the development of free apps with the ad benefits. To render advertising contents into an app, the developers typically register with a third-party mobile ad SDK provided by ad networks, such as AdMob~\cite{admob}, MoPub~\cite{mopub}, InMobi~\cite{inmobi}, etc. The ad networks grant developers with ad controls, such as displaying the ads with specific ad formats (depending on the platform, \textit{e.g.}, mobile or tablet). When achieving the ad-embedded page, the app sends an ad request to retrieve an ad to display from the ad network. The fetched ad content is then rendered on the end user's screen. The app developers are paid by the count of the ads displayed (\textit{e.g.}, impressions).

Generally, to maximize the ad revenue, the app developers would choose the ad networks that provide the highest payments. The ad payment strategies provided are indicated by two factors, fill rate and eCPM~\cite{goodnetwork}. Here, the fill rate represents the ratio of how many times an application requests to show an advertisement and how many times it actually shows up, and eCPM (``effective cost per mile'', also called RPM) describes the ad revenue generated per 1,000 impressions. The calculations are defined as the following.

\begin{equation}
Fill\, Rate = \frac{\# Impressions}{\# Ad\, Requests}
\end{equation}
\begin{equation}
eCPM = \frac{Total\, Earnings}{\# Impressions} \times 1000
\end{equation}

However, delivering a poor user experience can greatly impact the total ad revenue~\cite{calculator, goodnetwork}. Rendering too many ads, or using too big ads, interfere the users' feelings, and thereby reduce the user retention. The discontented users may post unfavorable reviews to express their complaints, which could be referred by the future potential users. This vicious spiral can ultimately lead to a small user volume of the app. Suppose average daily user sessions, average minutes per session, and ad impressions per minutes are $N_{user}$, $N_{min}$, and $N_{ad}$, respectively. The ultimate total ad revenue can be defined as

\begin{equation}
Ad\, Revenue = \frac{N_{user}\times N_{min}\times N_{ad}}{1000} \times eCPM \times Fill\, Rate.
\end{equation} 

With a poor user base, the ads could not benefit the developers with great revenue, even if the ad network offers high eCPM and fill rate. Therefore, a good user experience serves as the hinge of the whole profit process. In this paper, we are devoted to inspecting the ad costs from users' perspective. 
\section{Overall Framework}\label{sec:framework}
\begin{figure}[h]
    \centering
    \includegraphics[width=0.4 \textwidth]{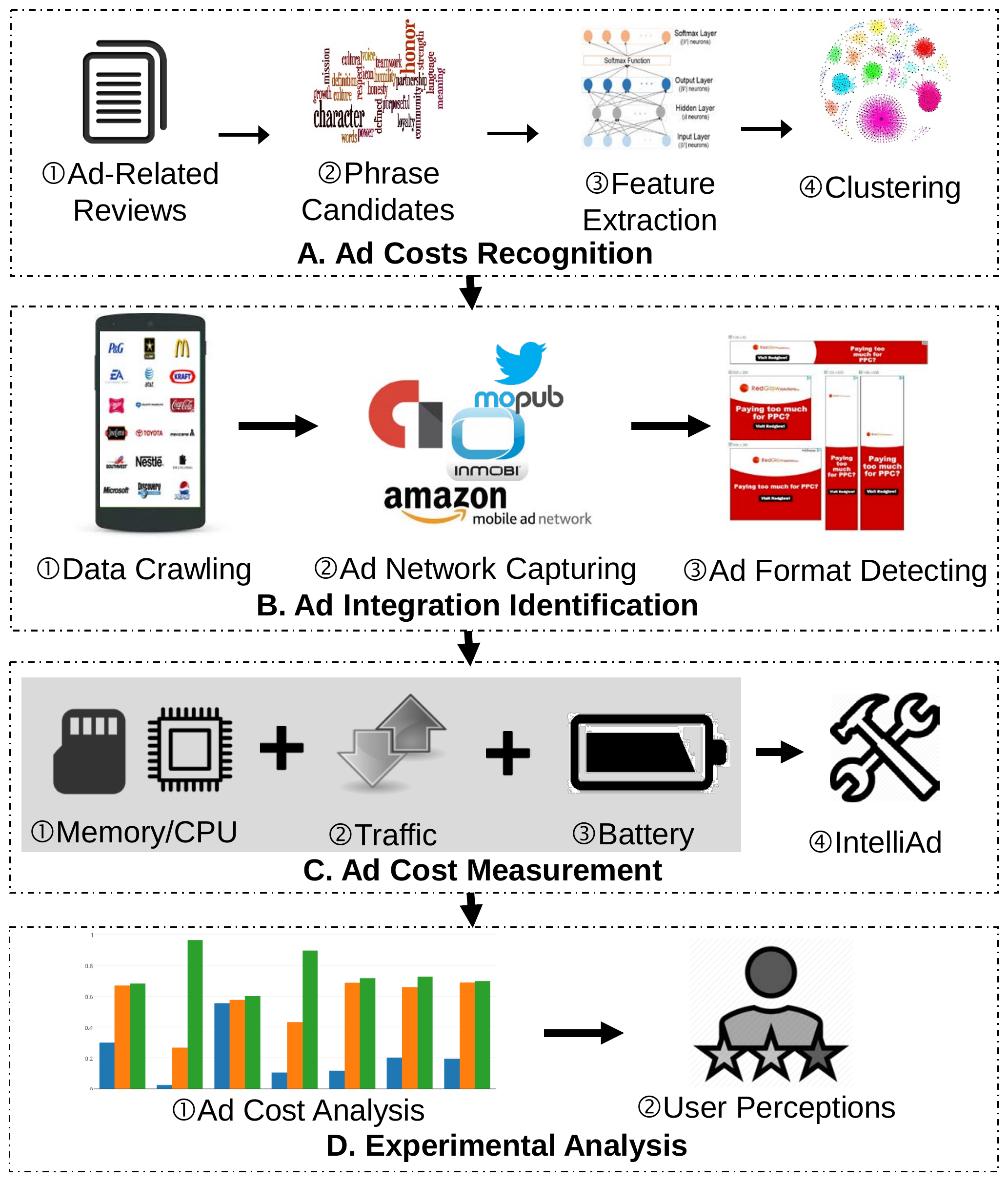}
    \caption{The Overall Framework of the Experimental Study.}
\label{fig:framework}
\end{figure}

The overall framework of the experimental study is outlined in Figure~\ref{fig:framework}, including following phases: ad costs recognition, ad integration identification, ad cost measurement, and experimental analysis.

We first recognize the ad-related user concerns from user reviews (Section~\ref{sec:cost}), and regard them as the ad costs to explore. Different from the previous studies, which are app-based, we measure the ad costs based on ad integration schemes. We suppose that ad embedding is necessary for the free app development, and verify whether different schemes generate different ad costs.

Generally, the ad integration schemes are the factors that the developers are required to determine for rendering ads into apps, including the ad network and the corresponding ad format. Ad formats define the displayed ad sizes (depending on the platform, \textit{e.g.}, mobile or tablet) or types (\textit{e.g.}, image or video). To obtain the commonly used ad integration schemes, we conduct an experimental study on 104 popular apps (listed in Table~\ref{tab:appdata}) of 19 categories from Google Play. We capture both the ad networks appended and the ad formats defined in the apps using static analysis (Section~\ref{sec:implementation}). Meanwhile, we obtain the number of ads, one type of ad costs, of each scheme. Then we design a tool named $\mathsf{IntelliAd}$ to automatically measure and profile the other three types of ad costs of each scheme (Section~\ref{sec:intelliad}). The three measured costs are memory/CPU overhead, traffic usage, and battery consumption. We then analyze and discuss the measurement results of different schemes regarding the four types of ad costs (Section~\ref{subsec:analysis}). To provide the developers with deeper insights on user perceptions, we further investigate the correlations between the measured ad costs and the corresponding user concerns (Section~\ref{subsec:users}). The experimental results demonstrate that users indeed care about several types of ad costs while ignoring the others.

\section{Ad Costs Recognition}\label{sec:cost}
App reviews have been considered as the descriptions of user perceptions~\cite{khalid2014mobile, kong2015autoreb}, and utilized to discover app issues during the app development~\cite{vu2015mining,gu2015parts}. Since over 70\% of users find that automatically served in-app ads ``annoying''~\cite{tapjoy} and they probably write negative reviews, analyzing the reviews can provide us inspirations on the user concerns. In this section, we first illustrate the method we have employed to extract the user-concerned ad costs, and then present the results.

\subsection{Analyzing User Reviews}
To comprehend the ad-related user concerns, we conduct an automatic review analysis on 399,5246 real app reviews. The reviews utilized are crawled from Google Play for the period between September 2014 and March 2016. From the review collection, we extract 19,579 (0.5\%) reviews that explicitly describe some facets about ads (regular expression = \textit{ad/advert*}). We then leverage the tool $\mathsf{PAID}$~\cite{paid2015cygao,paid} to extract the user concerns. We employ the tool since it has been demonstrated to be effective in observing app issues from user reviews. The ultimate issues are presented in phrases instead of words for better understanding. The whole analysis process is depicted in Step A of Figure~\ref{fig:framework}.


First, we obtain all the phrase candidates that could possibly indicate the user concerns from the ad-related reviews. Next, we employ continuous bag-of-word (CBOW) model, one flavor of word2vec model, to convert each phrase into a vector during the feature extraction. The outputs of word2vec are then fed into the clustering step to group the phrases into several topics. Finally, we analyze the phrases of each cluster to recognize the user concerns relevant to ads.

\subsection{Discovered User Concerns}
We employ k-means for the clustering step and set $k=4$. Table~\ref{tab:adkmeans} illustrates the extracted user concerns represented in phrases.
\vspace{-0.2cm}

\begin{table}[h]
\center
\small
   \caption{Selected Top Phrases of Each Cluster. The highlights indicate the ads costs concerned by users.}
    \label{tab:adkmeans}
    \begin{tabular}{|c||m{1.1cm}|m{1.8cm}|m{1.4cm}|m{1.2cm}|}
    		\hline
    		Topic & Topic 1 & Topic 2 & Topic 3 & Topic 4  \\
    		\hline
    		\multirow{2}{*}{1} & \hl{slow} & \hl{battery} & \hl{premium} & pls \\
    		 & \hl{loading} &  \hl{drain} & \hl{version} & remove \\
    		\hline
    		\multirow{2}{*}{2} & free & \hl{internal} & user & \hl{without} \\
    		& trial &  \hl{memory} & experience & \hl{wifi} \\
    		\hline
    		\multirow{2}{*}{3} & force & \underline{tried} & \hl{much} & becomes \\
    		& closing & \underline{uninstalling} & \hl{memory} & laggy \\
    		\hline
    		\multirow{2}{*}{4} & new & \hl{memory} & home & already \\
    		& layout & \hl{hog} & screen & paying \\
    		\hline
    		\multirow{2}{*}{5} & \hl{many} & \hl{battery} & dark & get \\
    		& \hl{ad} & \hl{consumption} & theme & annoying \\
    		\hline
    \end{tabular}
\end{table}

As Table~\ref{tab:adkmeans} shown, users complain about the \textit{annoying} features (the highlighted phrases) of ads through the reviews. When users feel intolerable of the ad displaying, they would uninstall the apps (``tried uninstalling'' in Topic 2) or change to a premium version (``premium version'' in Topic 3). Overall, we discovered that 0.3\% users clearly express ideas about ``uninstalling'', and more than 16.0\% of users describe ads as ``annoying''. We can estimate that the developers miss the ad benefits produced by the impressions from these users. We call the \textit{annoying} ad features as \textit{ad costs}.

To measure the ad costs more specifically, we classify them into the following four types according to Table~\ref{tab:adkmeans}. For each type of ad costs, we examine the corresponding reviews to ensure that the costs complained are indeed influenced by ads, not by the host apps.

\begin{compactitem}
\item \textbf{Number of ads} (\textit{e.g.}, ``many ad'' in Topic 1). For example, one user of app \textit{com.jb.zcamera} complains that ``So many ads and I paid money for the ad block and new filters and nothing happened''.
\item \textbf{Memory/CPU utilization} (\textit{e.g.}, ``memory hog'' in Topic 2). For example, one review of app \par
\noindent \textit{com.android.chrome} describes that ``Memory hog and need to add an exit button and ad blocker''.
\item \textbf{Traffic Usage} (\textit{e.g.}, ``without wifi'' in Topic 4). For example, one review of app \textit{com.facebook.katana} states that ``With how little use the phone without WiFi, used 400MB of data rate, opening it only once. And all notifications that arrive are you just advertising. Uninstalled''. 
\item \textbf{Battery Consumption} (\textit{e.g.}, ``battery drain'' in Topic 2). For example, another user of app \textit{com.android.chrome} expresses that ``More ads increase more battery consumption. Settings are fake''.
\end{compactitem}

These four types of ad costs are considered as the user concerns regarding the in-app ads, and are employed as metrics to evaluate the impact of the ad integration.
\section{Ad Integration Identification}\label{sec:implementation}
Ad integration identification aims at detecting the utilized ad schemes (\textit{i.e.}, ad networks and ad formats) in the apps. To invoke an ad API provided by the ad network, the developers must instantiate the ad rendering class (\textit{e.g.}, com.google.android.gms.ads.AdView) in the java code, illustrated in Figure~\ref{fig:adexample}. We collect the ad classes of the most popular 20 ad networks (ranked by AppBrain~\cite{adlibrary}), and reverse engineer all the Android apps by employing \textit{Apktool}~\cite{apktool}. Since the top 20 ad networks occupy over 70\% of the mobile advertising market, these ad networks are representative and considered preferred by the developers. If one class of the ad network is called in the decomplied code, we then regard that the corresponding ad SDK is integrated into the app. 

\begin{figure}[h]
    \centering
    \includegraphics[width=0.35 \textwidth]{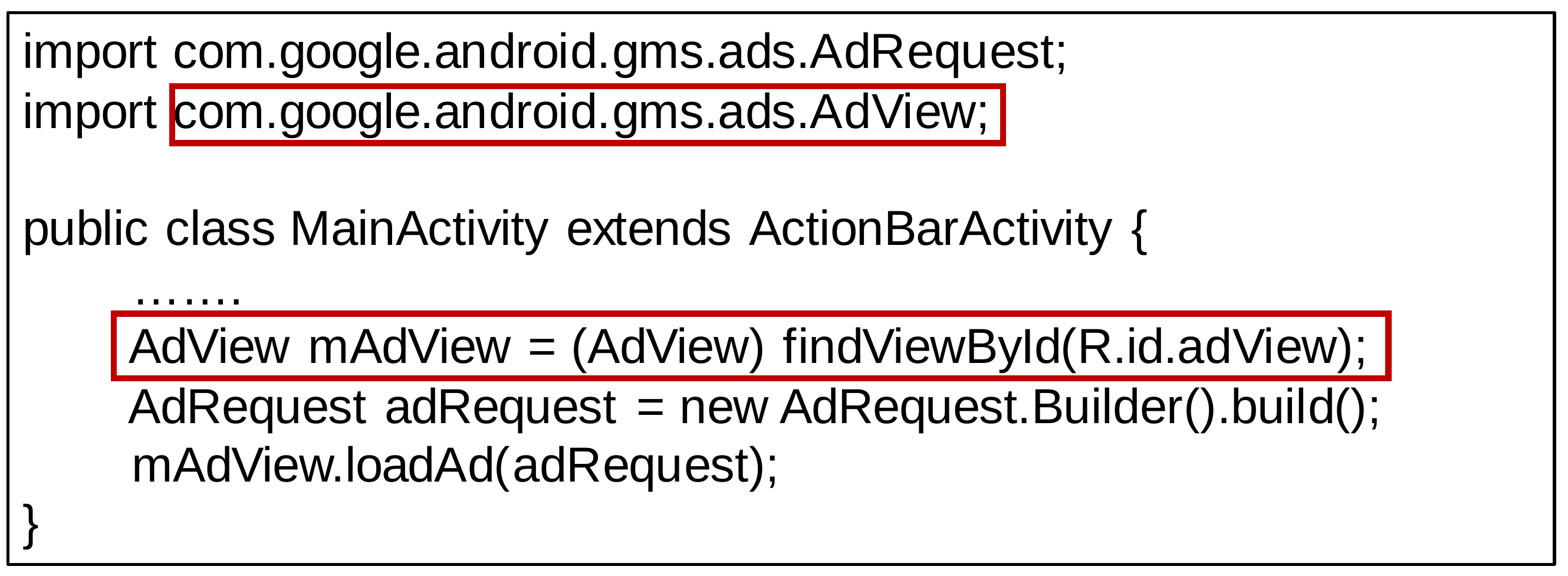}
    \caption{Code Snippet for Rendering AdMob Ad in the App.}
\label{fig:adexample}
\end{figure}

\begin{figure}[h]
    \centering
    \includegraphics[width=0.45 \textwidth]{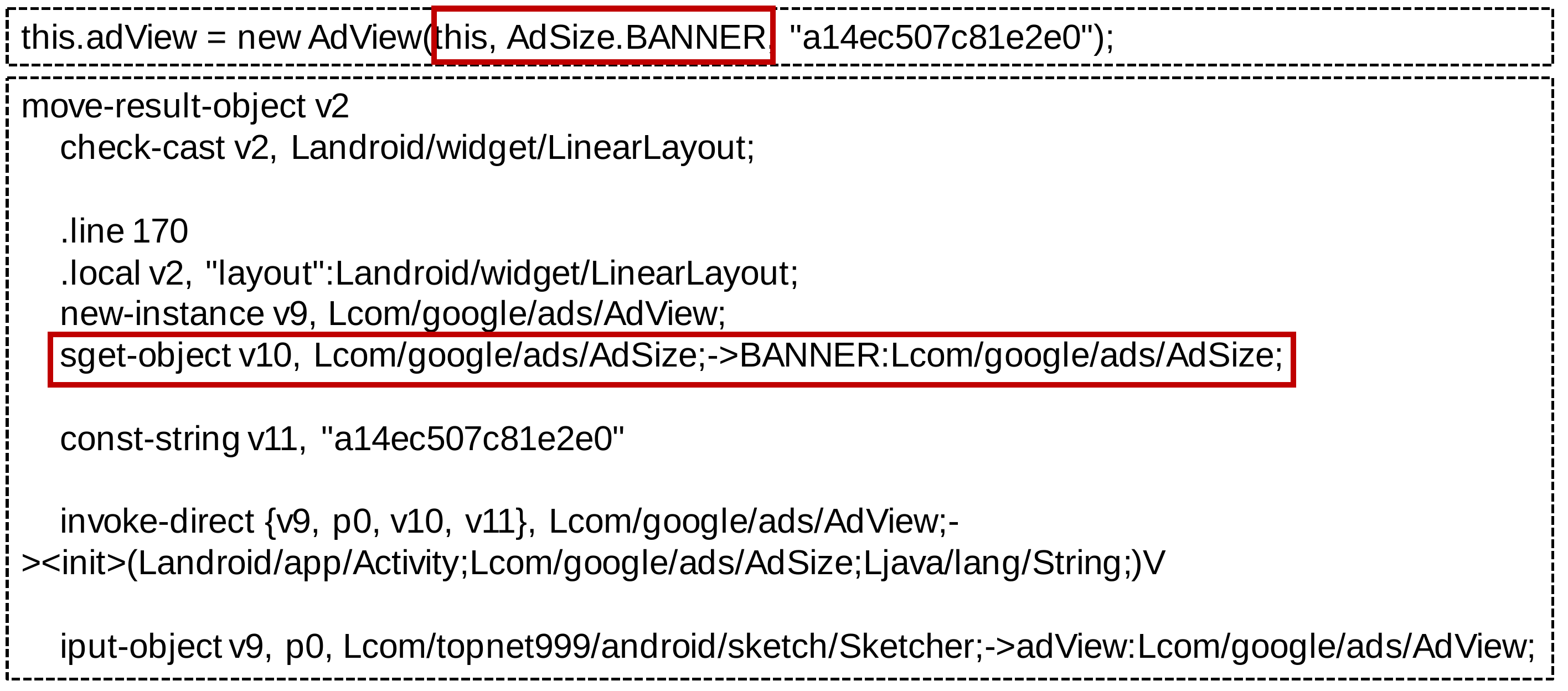}
    \caption{Code Snippets of the Original Source Code (top) and the Corresponding Decompiled Code (below).}
\label{fig:smali}
\end{figure}

Based on the discovered ad networks, we explore the corresponding ad formats defined by the developers. Generally, ad format can be categorized into three types, \textit{i.e.}, banner ads, interstitials, and video ads~\cite{adformat}. The definitions of ad formats mainly aim at rendering ads in appropriate sizes for different platforms, such as mobile and tablet. For example, the size of the banner ad on the mobile end can be defined as 320$\times$50, while on the tablet it is better to be rendered as 728$\times$90. Interstitials are full-page ads that appear in the app at natural breaks or transition points. Generally, the ad format can be declared in two ways, \textit{i.e.}, programmatically determined in the java code, or defined in the \textit{layout} file. Fig.~\ref{fig:smali} depicts the code snippet representing the original declaration based on the first method and the corresponding decomplied form. By static analysis, we extract the ad format - BANNER. Similarly, for the declaration in the second way, we detect the ad layout id from the decomplied code and then recognize the corresponding format from the \textit{layout} file. Finally, all the ad formats of the embedded ads are obtained. Meanwhile, we capture the number of ads incorporated in each app. The identified ad integration schemes are illustrated in Table~\ref{tab:result}.

\section{Ad Cost Measurement}\label{sec:intelliad}
In this part, we elaborate the measurement method for each type of ad cost, based on which a tool named $\mathsf{IntelliAd}$ is then designed to automatically measure and profile the ad costs for each ad integration scheme.

\subsection{Ad Cost Metrics}\label{sec:metrics}
The ad costs to measure are the user concerns summarized in Section~\ref{sec:cost} except the number of ads, which we have already obtained in Section~\ref{sec:implementation}. They are memory/CPU overhead, traffic usage, and battery consumption. The detailed metrics and measurement methods for each type of ad cost are elaborated in the following.

\subsubsection{Memory/CPU Overhead}
We utilize three runtime metrics to determine the memory/CPU overhead, \textit{i.e.}, \textbf{memory consumed}, \textbf{CPU utilization}, and \textbf{number of threads}.

Much consumption on the resources leads users to experience lag. For example, memory management in Android enables the system to allocate the precious resource. When the memory becomes constrained, the system slows down dramatically~\cite{ram}. The overall busyness of the system can be quantified by the CPU utilization~\cite{cputime}. We suppose that different ad SDKs manage the ad lifecycle differently, and observe their memory and CPU expended during the runtime. The above two metrics are obtained by employing a standard tool \textit{top} in Android. The \textit{top} tool monitors the portion of memory consumed (\textit{i.e.}, Resident Set Size or RSS) and the CPU occupancy rate of a process in real time. We run the tool in one second interval, and compute the average RSS value and CPU utilization for the subsequent analysis.

In addition, the app may also appear to hang, when the UI thread (also called the ``main thread'') performs intensive work. Since rendering ads in the user's interface involves the implementation of the UI thread, we also consider the number of threads as a metric of the memory/CPU overhead. The metric is evaluated through reading the \textit{/proc/pid/stat} file during the app runtime.

\subsubsection{Traffic Usage}
As Figure~\ref{fig:adflow} illustrates, the ad SDKs embedded in the mobile apps send requests to and fetch contents from ad networks for ad rendering. The whole process requires data transmission, further leading to Internet access cost or battery drainage~\cite{yoon2012appscope} for end users.

Different ad networks and ad formats (\textit{e.g.}, banner ads, interstitials, and video ads) can influence the traffic usage~\cite{mohan2013prefetching}. However, the developers lack visibility and attention regarding how the network consumption varies with different networks and formats. In the paper, we measure this type of ad cost by two metrics, namely \textbf{total bytes} and \textbf{number of packets}. Total bytes of data include the bytes sent and received by the app during the runtime. Similarly, the number of packets also considers both the network packets sent and received by the app. We utilize the typical tool \textit{tcpdump} to estimate the two metrics. The tool is started to run when the app is launched, and records all the data access information. The average values are regarded as the estimation of the ad cost.

\subsubsection{Battery Consumption}
Battery power is a limited resource on mobile devices. Constantly synchronizing and pushing alerts, such as ads, can hog battery life. To measure the battery consumption of each ad integration scheme, we leverage an effective measurement framework $\mathsf{AppScope}$~\cite{yoon2012appscope}. $\mathsf{AppScope}$ comprises five components (\textit{i.e.}, CPU, LCD, WiFi, cellular, and GPS). Since GPS and cellular are switched off during the experiments, and also the LCD settings are consistent in the course, we exclude these three factors when approximating the power consumed. We choose WiFi for the network connection.

To evaluate the battery consumed by WiFi, we capture the packet rate $p$ using~\textit{tcpdump}, which indicates the number of packets collected in one second interval. The battery drainage is estimated by the following equation, where $t$ denotes the threshold, and $\beta$ represents the coefficients.

\begin{equation}
P^{WiFi} = 
\begin{cases}
  \beta_{l}^{WiFi}\times p + \beta_{l}^{base}, & \text{if}\ p \leq t \\
  \beta_{h}^{WiFi}\times p + \beta_{h}^{base}, & \text{if}\ p > t
\end{cases}
\end{equation}

Towards estimating the battery drained by CPU, we obtain the average runtime CPU frequency through the utility\par
\noindent \textit{/sys/devices/system/cpu/cpu0/cpufreq/scaling\_cur\_freq}, which records the real-time speed of CPU. We denote the average CPU frequency as $freq$. Then the power consumed by CPU is defined by

\begin{equation}
P^{CPU} = \beta_{freq}^{CPU}\times u + \beta_{freq}^{idle},
\end{equation}

\noindent where $u$ denotes the average CPU utilization we have captured using \textit{top}, $0\% \leq u \leq 100\%$. We compute the coefficient, $\beta$, using the linear regression model based on the dataset provided by~\cite{yoon2012appscope}. The average standard deviations for $\beta_{freq}^{CPU}$ and $\beta_{freq}^{idle}$ are 9.21 and 1.84, respectively. Since $u$ ranges from 0 to 1, the deviations are acceptable for battery measurement~\cite{yoon2012appscope}. Finally, the average power consumption is determined by

\begin{equation}
P = P^{WiFi} + P^{CPU},
\end{equation}

\noindent which combines the battery drainage on both components (\textit{i.e.}, WiFi, and CPU) by addition.

\subsection{IntelliAd}\label{subsec:study}
The mobile device we have used is the LG Nexus 5 smartphone with a rooted Android 5.0.1 operating system. Our goal is to measure and profile ad costs automatically and consistently. In the following, we introduce the experimental settings of $\mathsf{IntelliAd}$ in detail.

\textbf{One Page, One Ad:} According to the mobile advertising policies~\cite{onepageads}, the number of banner ads on a single screen should not exceed one. For ad schemes with more than one banner ads, we render them in different activities. Since each scheme in Table~\ref{tab:result} involves at most two banner ads, we implement two activities in our experimental apps. For the schemes with two banner ads, we embed them separately into the apps.

\textbf{Ad Cost Separation:} We aim at collecting the costs produced by the integrated ads only.
To achieve this, we create a basic prototype app for incorporating different ad schemes. As Figure~\ref{fig:adexample} depicts, the prototype app (on the left) has three buttons, among which one navigating to another empty activity, and the other two are fake buttons. The fake buttons are utilized to render the interstitial ads in the schemes. The right screenshot presents an experimental app with MoPub banner ad rendered. By clicking the three buttons from top to bottom, the app screen displays MoPub interstitial ad, Amazon interstitial ad, and Amazon banner ad, respectively. Both the prototype app and the ad-embedded apps are measured during the experiment. The costs of the ad scheme are calculated by subtracting the costs of the prototype app from the measured costs of the corresponding ad-integrated app. In this way, the costs of each ad integration scheme are determined.

\begin{figure}[h]
    \centering
    \includegraphics[width=0.4 \textwidth]{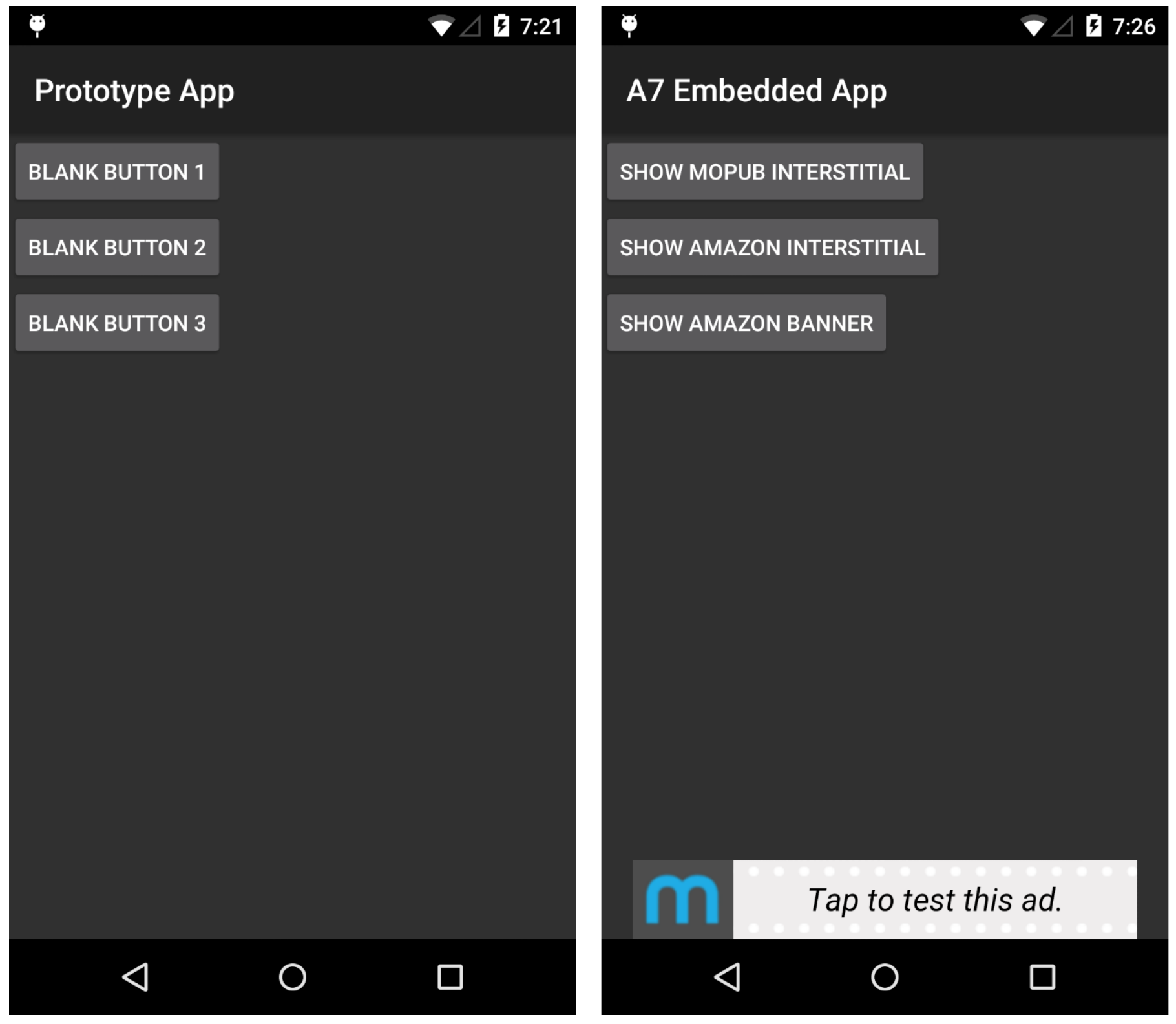}
    \caption{Prototype App and One Experimental Ad-Embedded App (A7 in Table~\ref{tab:result}).}
\label{fig:adexample}
\end{figure}

\textbf{Ad Display Control:} To automatically measure the costs and guarantee the ad display duration to be consistent, we leverage a dynamic analysis framework $\mathsf{AppsPlayground}$~\cite{rastogi2013appsplayground}. We modify the framework in two aspects to measure the ad costs more accurately and comprehensively. First, we adjust the execution rules. For the UI widget with ad rendering (\textit{e.g.}, the class name of the widget ending with ``View'', ``WebView'', or ``FrameLayout''), we skip the click operation. This is to make sure that the ad can only be displayed, and not clicked. Secondly, we define the interval between two operations (\textit{e.g.}, click) to be 20 seconds. Thus, the ad display durations are identical. 

\textbf{Measuring Frequency Setting:} When the app is launched, \textit{tcpdump} and \textit{top} are started to record the information about the traffic usage and the memory/CPU overhead. The \textit{top} tool is set to run at one second interval. Between two operations (20s), the number of threads and the CPU frequency are captured by reading the system files every 0.04s. 

\textbf{Mitigating Background Noise:} To mitigate the background noise, for each app version, we first restore the system environment to its original state. Then we install the app and start its execution. Furthermore, each ad integration scheme is measured four times during the experiments to minimize the noise.

\textbf{Source of Ad:} As the mobile advertising policy declares, clicking on the live ads is forbidden during development and testing. All the ads examined are defined as test mode.

\section{Experimental Study}\label{sec:experiment}
In this section, we elaborate the experiments conducted on 104 real apps belonging to 19 categories (listed in Table~\ref{tab:appdata}) from Google Play. The apps are top 100 apps of each category according to AndroidDrawer~\cite{androiddrawer}. With multiple categories and enough popular apps, we can obtain the commonly used ad integration schemes comprehensively. Table~\ref{tab:result} summarizes the ad embedding conditions of these apps, including the ad networks and the corresponding ad formats.

\begin{table}[ht]
\small
\center
   \caption{Subject Applications.}
    \label{tab:appdata}
    \begin{tabular}{|m{3cm}||c|}
    		\hline
    		Category & \# App \\
    		\hline
    		Business & 4 \\
    		\hline
    		Books \& References & 6 \\
    		\hline
    		Comics & 3 \\
    		\hline
    		Education & 2\\
    		\hline
    		Finance & 2 \\
    		\hline
    		Health \& Fitness & 15 \\
    		\hline
    		Lifestyle & 3 \\
    		\hline
    		Media \& Video & 6 \\
    		\hline
    		Medical & 2 \\
    		\hline
    		Music \& Audio & 6 \\
    		\hline
    		News \& Magazines & 9 \\
    		\hline
    		Personalization & 6 \\
    		\hline
    		Sports & 1 \\
    		\hline
    		Tools & 1 \\
    		\hline
    		Productivity & 20 \\
    		\hline
    		Social & 6 \\
    		\hline
    		Shopping & 1 \\
    		\hline
    		Photography & 7 \\
    		\hline
    		Weather & 4 \\
    		\hline
    \end{tabular}
\end{table}

\begin{table*}[ht]
\small
\center
   \caption{Ad Integration Scheme Summary.}
    \label{tab:result}
    \begin{threeparttable}
    \begin{tabular}{c|c|c|c|c|c|c|c|c}
    		\hline
    		\multirow{3}{*}{ID} & \multirow{3}{*}{Ratio (\%)} & \multicolumn{5}{c|}{Ad Integration Scheme} & \multirow{3}{*}{\# Review} & \multirow{3}{*}{Avg. Rating} \\ \cline{3-7}
    		
    		& & \multirow{2}{*}{Ad Network} & \multicolumn{4}{c|}{Ad Format} &  \\ \cline{4-7}
    		& & & Banner\tnote{1} & Smart\_Banner\tnote{2} & Full\_Banner\tnote{3} & Interstitial\tnote{4} & & \\ 		
    		\hline
    		A1 & 42.3 & AdMob & \checkmark & & & & 62,918 & 4.12 \\
    		\hline
    		A2 & 13.5 & AdMob & \checkmark & & & \checkmark & 6,020 & 3.62 \\
    		\hline
    		A3 & 12.5 & AdMob &  & \checkmark & & & 20,998 & 4.34 \\
    		\hline
    		A4 & 6.7 & AdMob & & \checkmark & & \checkmark & 6,478 & 4.14 \\
    		\hline
    		A5 & 4.8 & Amazon & \checkmark & & & & 10,720 & 3.50 \\
    		\hline
    		A6 & 4.8 & MoPub & \checkmark & & & \checkmark & 4,720 & 3.85 \\
    		\hline
    		\multirow{2}{*}{A7} & \multirow{2}{*}{3.8} & MoPub & \checkmark & & & \checkmark & \multirow{2}{*}{4,871} & \multirow{2}{*}{4.50} \\
    		& & Amazon & \checkmark & & & \checkmark & & \\
    		\hline
    		A8 & 3.8 & AdMob & & & \checkmark & & 7,517 & 4.15 \\
    		\hline
    		A9 & 2.9 & MoPub & \checkmark & & & & 7,256 & 4.43 \\
    		\hline
    		A10 & 2.9 & AdMob & & & & \checkmark & 3,247 & 4.07 \\
    		\hline
    		\multirow{2}{*}{A11} & \multirow{2}{*}{1.0} & AdMob & & \checkmark & & & \multirow{2}{*}{1,960} & \multirow{2}{*}{4.70} \\
    		& & MoPub & \checkmark & & & & &  \\ 
		\hline
		\multirow{2}{*}{A12} & \multirow{2}{*}{1.0} & AdMob & & \checkmark & & & \multirow{2}{*}{1,960} & \multirow{2}{*}{4.70} \\
    	    & & InMobi & \checkmark & & & & & \\    		
    		\hline
    \end{tabular}
    \begin{tablenotes}
    \scriptsize
\item[1] ``Banner'' refers to the banner ad in standard size (320$\times$50). \\
\item[2] ``Smart\_Banner'' represents the format that can adjust depending on the portrait or landscape orientation of the device viewing ads. \\
\item[3] ``Full\_Banner'' indicates full-size banner ads (468$\times$60). \\
\item[4] ``Interstitial'' describes the ads that cover the interface of the host app.
\end{tablenotes}
    \end{threeparttable}
\end{table*}

We strive to explore the answers of the following questions: 1) Does the ad integration scheme impact the ad costs concerned by users? (Section~\ref{subsec:analysis}) 2) Can the measured cost values reflect the user perceptions? (Section~\ref{subsec:users})

\subsection{Ad Cost Analysis}\label{subsec:analysis}
In this section, we first analyze the collection of ad integration schemes obtained during the static analysis in detail (Section~\ref{subsub:static}). Then based on the statistics profiled during the app execution, we analyze the ad costs of each ad integration scheme, and explore whether different schemes exhibit clear distinctions on these costs (\textit{i.e.}, memory/CPU overhead, traffic usage, and battery consumption).

\subsubsection{Number of Ads}\label{subsub:static}
\begin{figure}[h]
\centering
%

    \includegraphics[width=0.4 \textwidth]{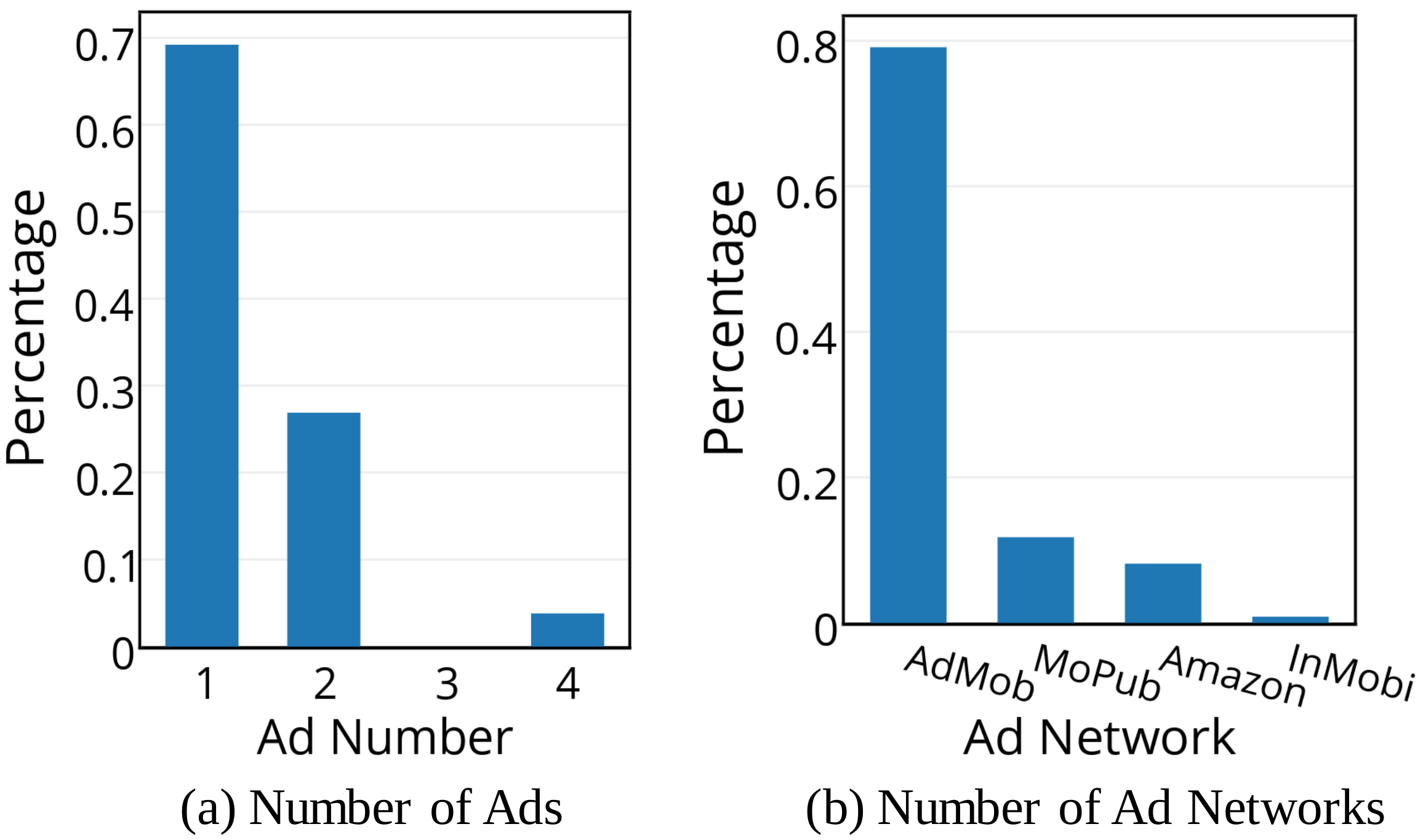}
    \caption{Overall View of the Subject Ad Integration Schemes.}
\label{fig:number}
\end{figure}

Based on the results of the ad integration identification (Section~\ref{sec:implementation}), we summarize the number of ads and ad networks of the subject apps in Figure~\ref{fig:number}. 

As Figure~\ref{fig:number} illustrates, the majority apps (69.2\%) are incorporated with only one ad, while the remaining (26.92\%) are mainly rendered with two ads. It is worth noting that some apps even display four ads. The behavior of incorporating so many ads may be driven by ad revenue. However, as we have discussed in Section~\ref{sec:cost}, users tend to complain about the ad numbers in the reviews. This indicates that embedding more ads cannot necessarily guarantee more earnings without keeping user retention. Thus, we consider that different ad schemes receive different user feedback from the point of ad quantity. 

For the ad SDK integrated, AdMob undoubtedly occupies the largest proportion (79.09\%), with MoPub followed after (11.82\%). InMobi only accounts for less than 1\% of the subjects. Referring to the popular ad libraries provided by AppBrain~\cite{adlibrary}, we suppose that the four ad networks captured are more widely employed by the developers, which illustrates that our obtained ad schemes are representative for in-app advertising.

\subsubsection{memory/CPU Overhead}
\begin{figure}[h]
    \centering
    \includegraphics[width=0.4 \textwidth]{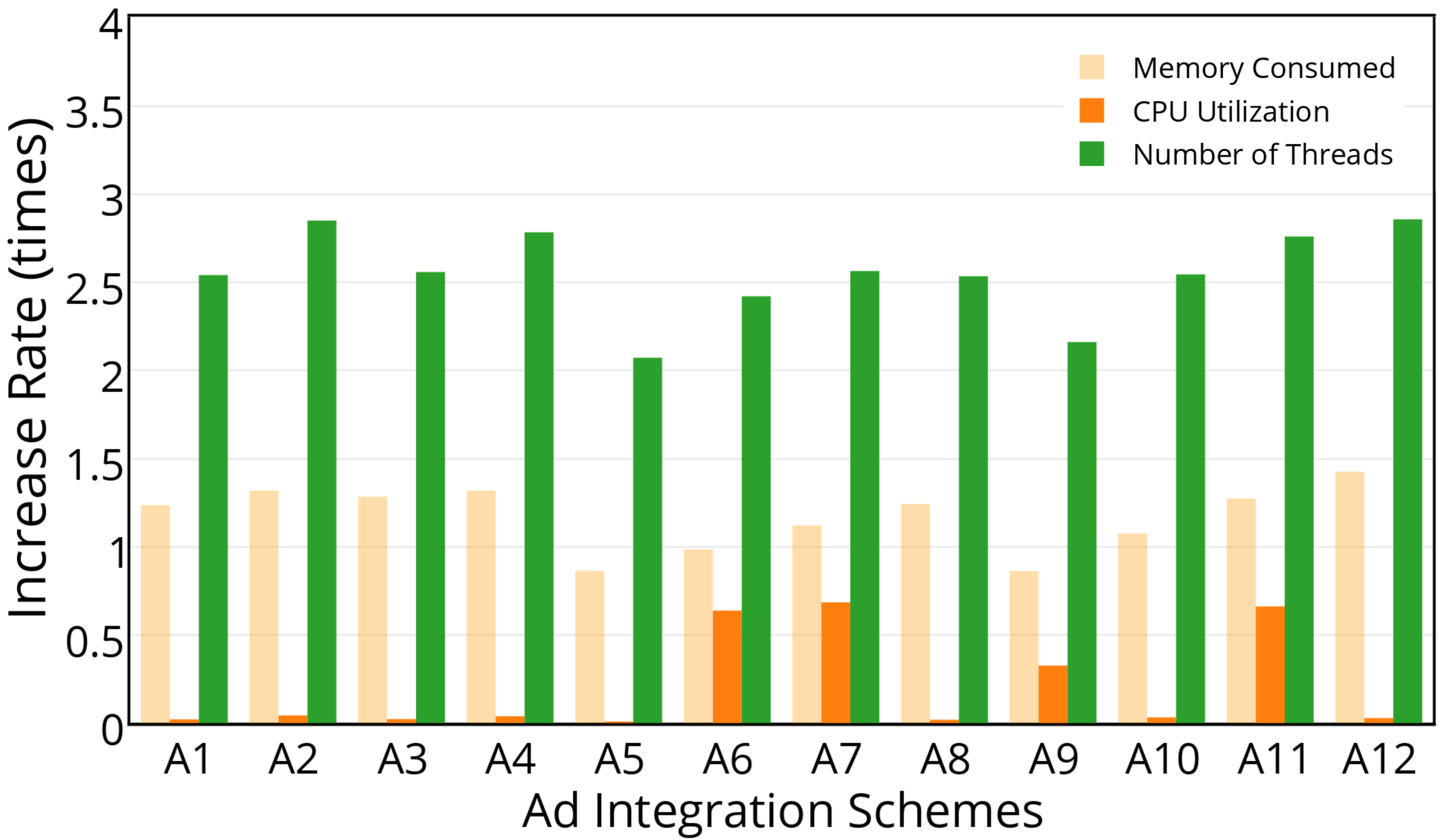}
    \caption{Increase Rate of Memory/CPU Utilization for Different Ad Integration Schemes.}
\label{fig:memo}
\end{figure}

The memory/CPU overhead is evaluated by three metrics, namely, memory consumed, CPU utilization, and number of threads. We calculate the increase rate of each metric with respect to the prototype version. Figure~\ref{fig:memo} illustrates the corresponding results for the 12 ad integration schemes.

As Figure~\ref{fig:memo} shows, the ad costs of all the ad integration schemes represent a growth trend in the three metrics. The average increase rates for the memory consumed and the number of threads are 1.17 times and 2.55 times, respectively, while CPU utilization just increases 0.21\% on average. However, the CPU utilization manifests the most obvious changes along with different ad schemes (avg. stdev 0.285), while the other two present relatively smaller differences (0.185 for memory consumed, and 0.248 for the number of threads). Thus, we employ the CPU utilization for the next analysis.

Examining the CPU utilization, we find that its high standard deviation is mainly influenced by A6, A7, A9, and A11. Taking a deeper look at these schemes, we discover that all of them include the incorporation of MoPub ad SDK. We suppose that such a high increase rate may be attributed to the MoPub SDK. Furthermore, the ad format influences the ad cost. For example, although A1 (Banner), A3 (Smart\_Banner), A8 (Full\_Banner), and A10 (Interstitial) are all rendered with AdMob only, they show different performance on the metrics. The increase rates of CPU utilization for these four schemes are 2.26\%, 2.42\%, 2.00\%, and 3.40\%, respectively. The interstitial ad demonstrates the highest ad cost. In addition, the results also vary with the number of ads embedded. For example, A7 with the most ads integrated presents the highest increase rate of all (68.60\%). In the experiment, the app with A7 integrated consumes 20.51\% CPU utilization. As we know, users may experience lag, when the CPU usage is above 70\%~\cite{cputime}. This indicates that the system may slow down with four such apps running simultaneously. Overall, we suppose that different ad integration schemes present different performance on the memory/CPU overhead, and could generate different user experience.

\subsubsection{Traffic Usage}
\begin{figure}[h]
    \centering
    \includegraphics[width=0.35 \textwidth]{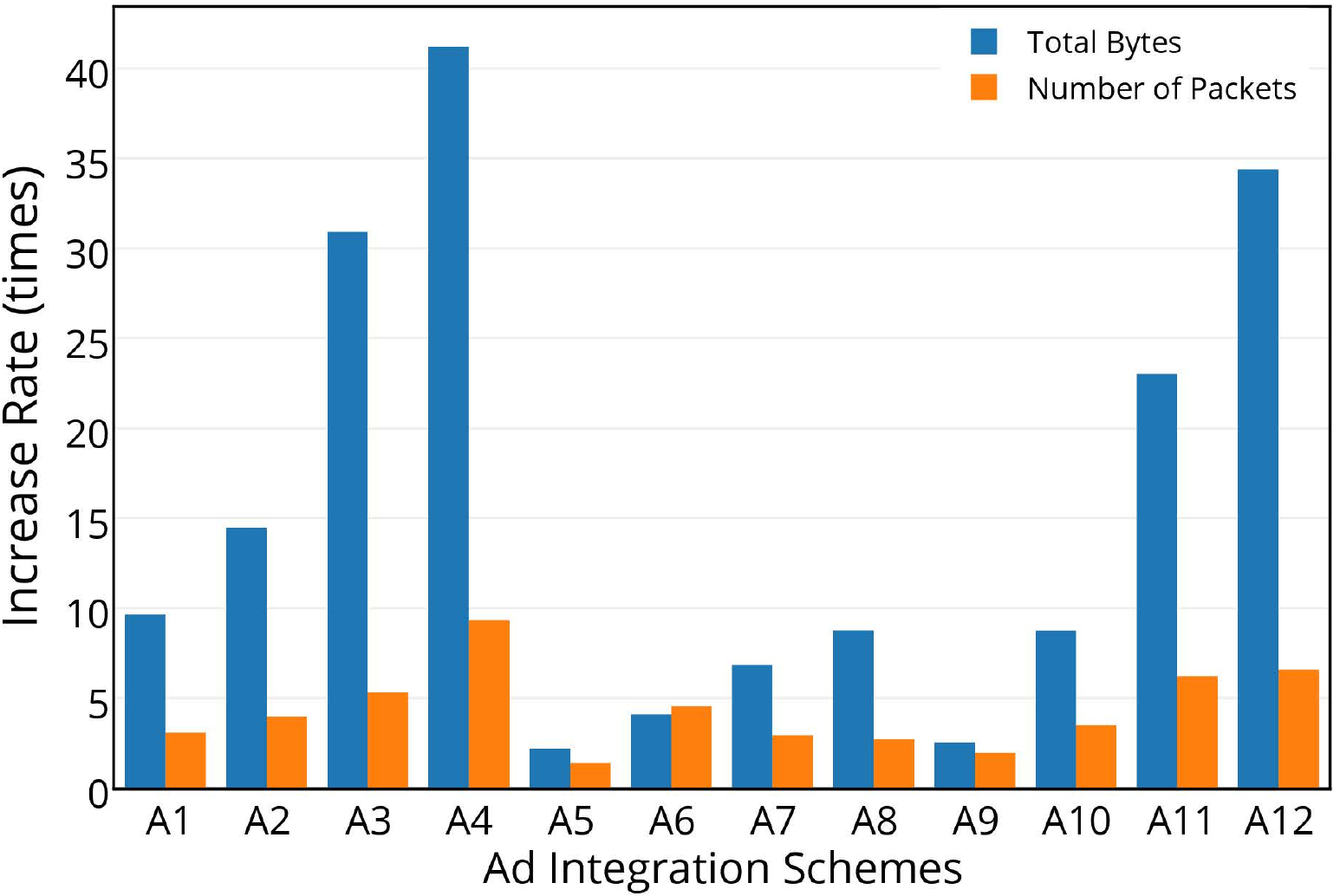}
    \caption{Increase Rate of Traffic Usage for Different Ad Integration Schemes.}
\label{fig:traffic}
\end{figure}

The traffic usage is estimated by two metrics, \textit{i.e.}, total bytes and number of packets. Figure~\ref{fig:traffic} depicts the growth rates of the metrics for each integration scheme regarding the prototype app.

As Figure~\ref{fig:traffic} illustrates, the total bytes and number of packets transmitted have increased simultaneously for all the ad-embedded versions. The average growth rates for both metrics are 15.56 times, and 4.29 times, respectively. Clearly, different ad schemes present different variations. For example, the apps with AdMob SDK integrated (\textit{e.g.}, A1, A2, A3, and A4) demonstrate more obvious increase trends than the other apps (\textit{e.g.}, A5, A6, and A7). A5 with only the Amazon banner ad embedded consumes the least network traffic (2.17 times for total bytes), while the other banner ads (A1 with AdMob Banner, and A9 with MoPub Banner) display slightly more network consumption (9.63 times, and 2.53 times, respectively). Furthermore, A4 with two ads embedded shows more traffic usage than A10. Integrated with ad SDKs that consume numbers of data bytes could spend users more money on Internet access. For example, suppose that the traffic consumed per user session is 896,926 bytes (same as the measuring result of A4), and a user of the major US carrier (AT\&T) selects the monthly data access plan \$25 for 5GB~\cite{atandt}. From this, we calculate that the user would spend \$0.0042 on ads during each execution of such an app. For other data access plan, the ad-related network charge would be more. Therefore, we conclude that the ad integration scheme impacts the traffic usage, and would further influence the users' interests.

\subsubsection{Battery Consumption}
\begin{figure}[h]
    \centering
    \includegraphics[width=0.35 \textwidth]{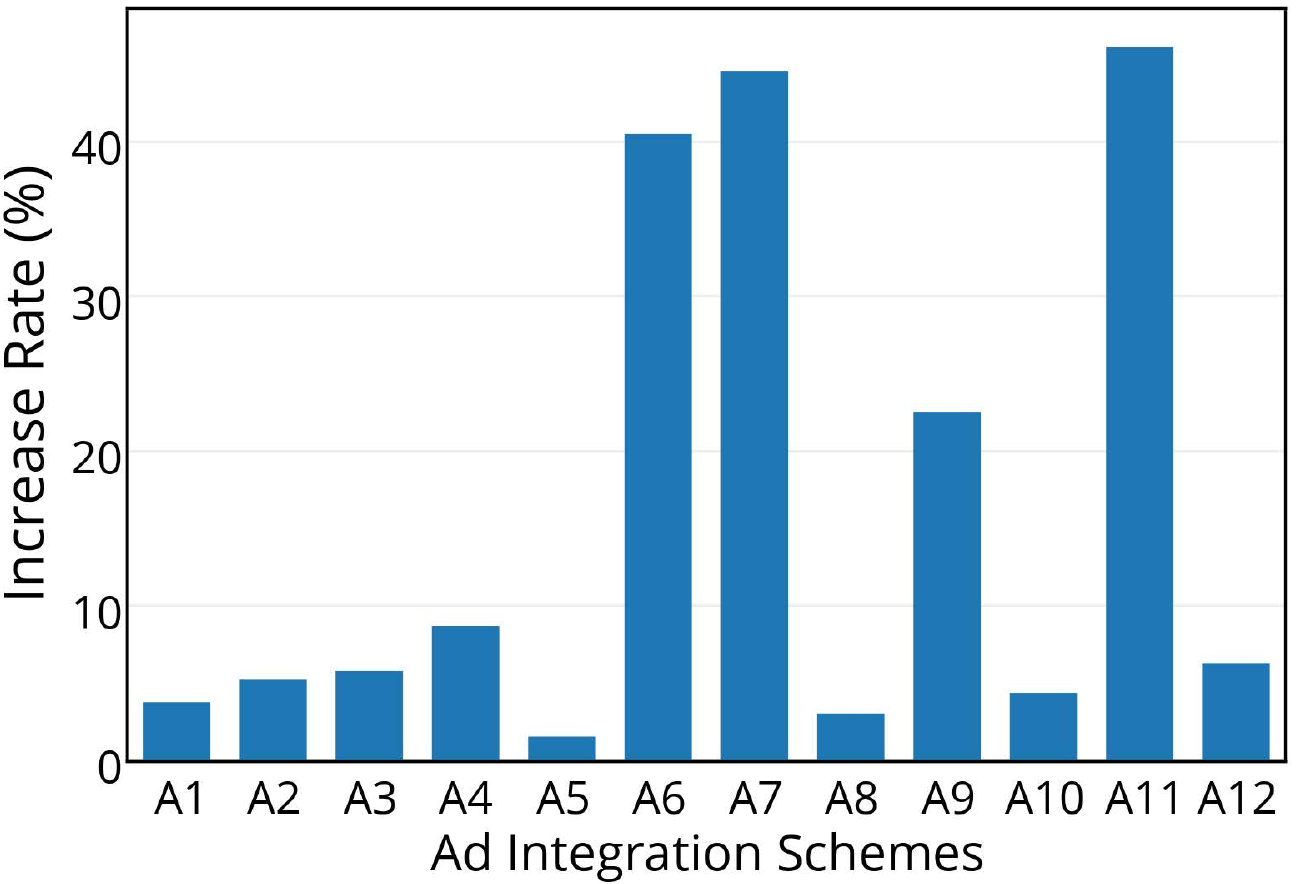}
    \caption{Increase Rate of Battery Consumption for Different Ad Integration Schemes.}
\label{fig:battery}
\end{figure}

The battery consumption is estimated based on $\mathsf{AppScope}$, illustrated in Section~\ref{sec:metrics}. The increase rates of different ad integration schemes are depicted in Figure~\ref{fig:battery}.

As Figure~\ref{fig:battery} shows, all the ad schemes generate certain battery drainage. The average increase rate of battery consumption for the 12 ad schemes is 16.03\%. Similar to the measuring results of memory/CPU overhead and traffic usage, different schemes present different performance on the battery consumed. For example, A6, A7, A9, and A11 display remarkable increase ratios than the other ad schemes (40.52\%, 44.55\%, 22.54\%, and 46.12\%, respectively). The high battery consumption may also be attributed to the incorporation of MoPub SDK. For the schemes involving only one ad (\textit{e.g.}, A1, A5, and A9), their performance also distinguishes, in which the Amazon banner ad (A5) shows the least cost (1.5\%), and the MoPub banner ad (A9) consumes the most battery. A7 with the most ads (four ads) integrated presents the highest power cost (44.55\% more). This indicates that for the smartphone (\textit{e.g.}, LG Nexus5) containing 2.2 hours of charge, the battery life would be drained to 1.2 hours with the ad displaying. Users must recharge the phones more often for offsetting the decreased energy cost, which would contribute to a poor user experience. Overall, we consider that different ad integration schemes exhibit different performance on battery consumption.

\subsubsection{Overview of Metrics}
\begin{figure}[h]
    \centering
    \includegraphics[width=0.3 \textwidth]{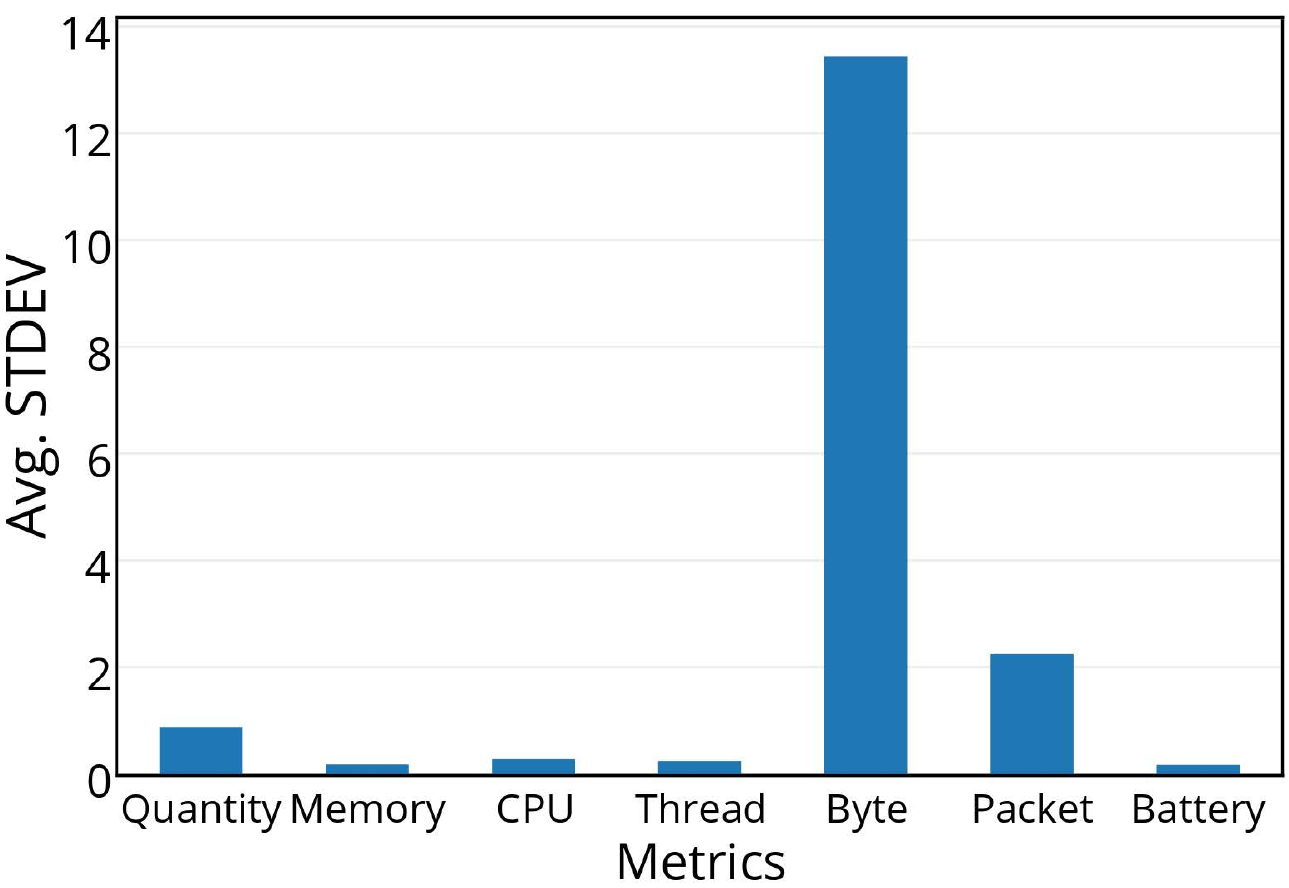}
    \caption{Average Standard Deviation for Different Types of Ad Costs.}
\label{fig:overall}
\end{figure}

Based on the distinct costs produced by different ad schemes, we explore which type of ad costs manifests the most significant variation. Figure~\ref{fig:overall} illustrates the average standard deviation for each type of cost regarding the metrics we have defined. The ``Quantity'' represents the number of ads. The next three metrics are utilized to measure the memory/CPU overhead. ``Byte'' and ``Packet'' indicate the total data bytes and number of packets, respectively. ``Battery'' refers to the battery consumption.

\vspace{-0.3cm}

From Figure~\ref{fig:overall}, we discover that CPU utilization and total bytes present the most salient variations in the corresponding ad costs, which means that ad integration can greatly affect the CPU performance and the data bytes transmitted. This observation is reasonable since rendering ads requires the CPU to execute the corresponding programs and also the network to request and fetch ads from the ad networks. Battery consumption displays the minimum difference among the schemes. Therefore, we suppose that the developers can detect the CPU utilization and the data bytes transferred to evaluate the ad integration scheme. Furthermore, the developers should consider the number of ads when embedding ads into apps.


\subsection{User Perceptions}\label{subsec:users}
Generally, the developers derive ad profits from the ads displayed (\textit{impressions}). Therefore, user retention serves as an essential role in the ultimate ad revenue. We have identified four types of ad costs concerned by users in Section~\ref{sec:cost}, and implemented the measurements on these costs in Section~\ref{sec:intelliad}. In this section, we aim at exploring whether the measured values match the user perceptions, \textit{i.e.}, whether there exists a gap between the experimental results and the user feedback. Here, the user reviews are regarded as the expressions of user perceptions.

We have collected 138,287 reviews of the subject apps from Google Play during April, 2016 (listed in Table~\ref{tab:result}). Each ad integration scheme has 11,524 reviews on average, which is large enough for review analysis~\cite{chen2014ar}. To capture the user perceptions on the four types of ad costs, we determine the particular keywords or key phrases corresponding to the ad costs. The keywords and key phrases are selected according to the process of discovering ad-related user concerns (Section~\ref{sec:cost}) and also the prior study~\cite{gui2015truth}, shown in Table~\ref{tab:keyword}. The reviews are considered to express the related costs if the corresponding keywords or key phrases are contained. Furthermore, we only consider the reviews with less than three stars to ensure that the users are dissatisfied with the ad costs.

\begin{table}
\small
\center
   \caption{Keywords and Key Phrases Corresponding to Each Type of Ad Cost.}
    \label{tab:keyword}
    \begin{tabular}{m{2cm}||m{5cm}}
    		\hline
    		Cost Type & Keywords  \\
    		\hline
    		\parbox[t]{2cm}{Number of \\ Ads} & many ad, much ad, free version, paid app, free app, lot of ad \\
    		\hline
    		\parbox[t]{2cm}{Memory/CPU \\ Utilization}& memory, slow, hang, ram, cpu, file, wait, laggy, lagging, delay, suspend\\
    		\hline
    		\parbox[t]{2cm}{Traffic \\ Usage} & bandwidth, wifi, network, data rate \\
    		\hline
    		\parbox[t]{2cm}{Battery \\ Consumption} & battery, drain, drainage, charge, recharge, power \\
    		\hline
    \end{tabular}
\end{table}

\begin{figure*}
    \centering
    \includegraphics[width=0.65 \textwidth]{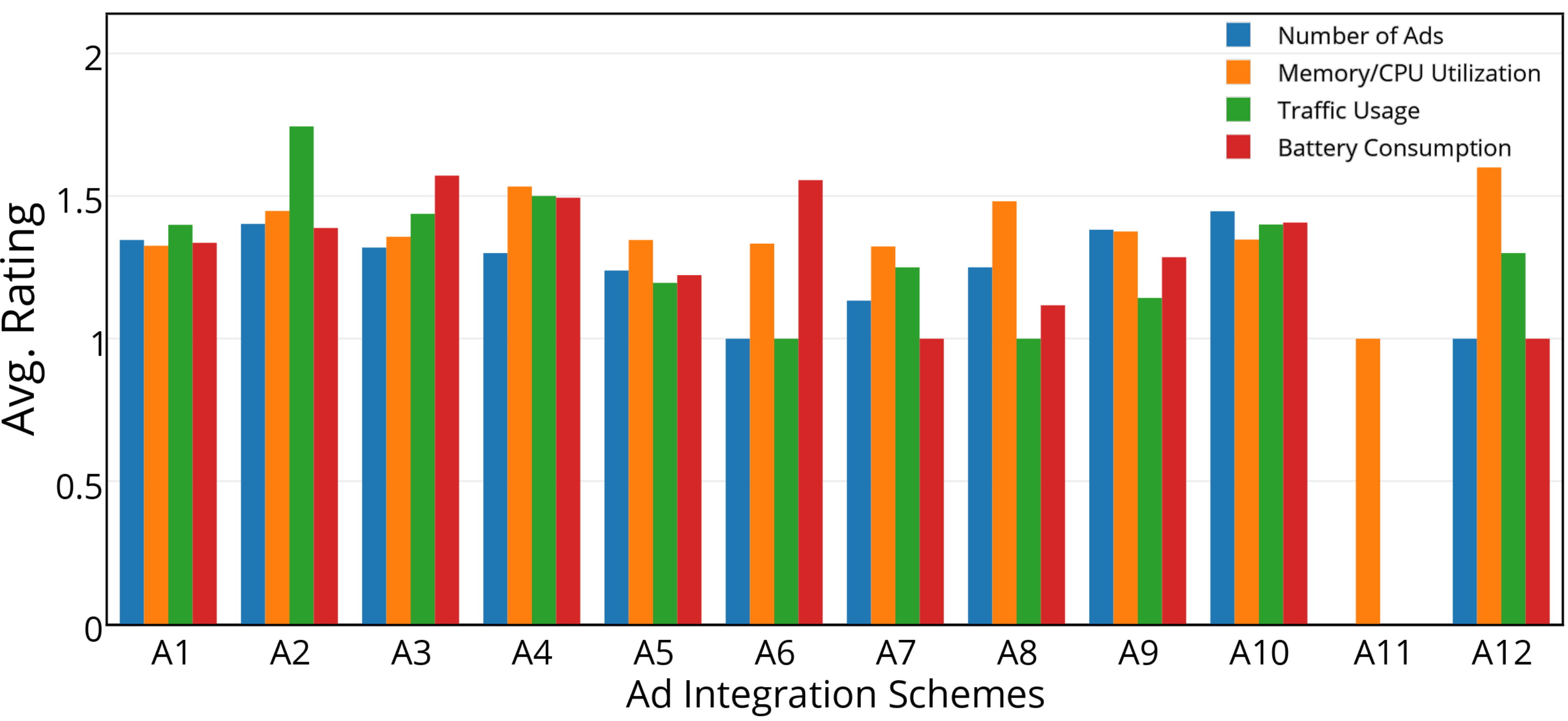}
    \caption{Average Rating of Each Ad Cost Type Described in the User Reviews Regarding the Ad Integration Schemes.}
\label{fig:review_data}
\end{figure*}

The average ratings of the ad costs for the schemes are presented in Figure~\ref{fig:review_data}. Regarding the number of ads, memory/CPU overhead, traffic usage, and battery consumption, the average ratings are 1.152, 1.373, 1.197, and 1.198, respectively. Combined with the previous measuring results for the ad schemes, we discover that their ad costs indeed reflect in the user reviews, explained in the following.

\noindent (1) \textbf{Number of ads.} A6, A7 and A12, with more than one ads embedded are rated lower by users (1.0, 1.3, and 1.0, respectively). We examine the app \textit{dk.boggie.madplan.android} belonging to A12 and find some reviews related to the ad quantity, \textit{e.g.}, one user complains that ``Use pro version still face too much ads''.

\noindent (2) \textbf{Memory/CPU overhead.} With a higher CPU utilization, A6, A7 and A11 receive lower user ratings about the ad cost than other schemes. A1 with larger memory consumption (7.0\% more than the average) displays more negative reviews about the memory overhead. We take one app \textit{com.rechild.advancedtaskkiller} using A1 for illustration. Some reviews describe that ``All it reliably do be pop up more ads and spawn process which consume even more of the phone resources'', and ``App be great but it freeze my phone with all the ads every time I do a kill from the widget''. This indicates that memory/CPU overhead can really spoil the user experience, even if the functions of the app are favorable.

\noindent (3) \textbf{Traffic Usage.} A12, presenting higher traffic usage in the experiments, also has a poorer user rating accordingly. With a slightly more data bytes transmitted, A7 is poorly scored. We take \textit{com.tinymission.dailycardioworkoutfree} for example. One review describes that ``Beware this app use leadbolt ad network which place ads in your notification bar in the background even if you aren't currently use this app''.

\noindent (4) \textbf{Battery Consumption.} Both A7 and A9 exhibit higher power drainage during the measurement, which are also reflected in the user ratings. We utilize one app\par
\noindent (\textit{com.tinymission.dailyabworkoutfree1}) embedded with A7 for explanation. One user states that ``Why do they want your location to drain your battery and send you even more ads''.

However, the gap between the measured values and the user perceptions is also obvious. For example, A2, A3 and A4, with the most traffic usage, are all relatively higher rated. To better comprehend the correlations between these aspects, we compute their Pearson correlations (listed in Table~\ref{tab:pearson}). We discover that the two aspects are negative correlated with respect to different types of ad costs except the traffic usage. This illustrates that higher ad costs can indeed result in lower user ratings. The memory/CPU overhead and battery consumption present the strongest correlations among all the cost types, which would indicate that users are more concerned with the system performance. However, the measured traffic usage, which we have demonstrated its obvious variations for different ad schemes, shows almost no correlations with the user perceptions. This may be attributed to the fact that WiFi has penetrated into people's daily life, leading to fewer concerns on traffic consumed. According to~\cite{whitepaper}, over 90\% users choose WiFi connections when using smartphones. The correlations also represent that all the ad costs are not strongly linearly correlated (less than 0.8) with the user ratings. We ascribe this to the influence of the hosts apps. Users may more concentrate on the characteristics of the host apps, but we aim at demonstrating the possible correlations between the ad costs and the user ratings here.

Therefore, we conclude that end users may not be sensitive about the network traffic. Overall, we identify that there exist both correlations and gaps between the measured values and the user perceptions. For integrating ads, we suggest the developers to consider the memory/CPU overhead, and also the battery consumption, to ensure the user experience.

\begin{table}
\small
\center
   \caption{Correlations Between the Measured Values and the User Perceptions (Average User Ratings).}
    \label{tab:pearson}
    \begin{threeparttable}
    \begin{tabular}{m{2cm}||M{3cm}}
    		\hline
    		Cost Type & Pearson Correlation  \\
    		\hline
    		\parbox[t]{2cm}{Number of \\ Ads} & -0.313 \\
    		\hline
    		\parbox[t]{2cm}{Memory/CPU \\ Overhead \tnote{1}}& -0.621\\
    		\hline
    		\parbox[t]{2cm}{Traffic \\ Usage\tnote{2}} & 0.081 \\
    		\hline
    		\parbox[t]{2cm}{Battery \\ Consumption} & -0.511 \\
    		\hline
    \end{tabular}
    \begin{tablenotes}
    \scriptsize
\item[1] Measured by the CPU utilization.\\
\item[2] Measured by the total data bytes.
\end{tablenotes}
    \end{threeparttable}
\end{table}

\vspace{-0.4cm}



\section{Discussion and Limitation}\label{sec:discussion}
In this section, we discuss the threats to validity of our study and illustrate the steps we have taken to mitigate the threats.

\textbf{External Validity:} Firstly, our experimental study and results are based on 104 real apps from Google Play, which represent an extremely small part of all the Android apps. We alleviate this threat by ensuring that all the subject apps are popular apps listed by AndroidDrawer and distributed in 19 different categories. Secondly, we implement and testify our framework on apps from Google Play. It is uncertain whether our methodology may be applicable for mobile ads in other stores (\textit{e.g.}, App Store and Amazon AppStore). However, since the ad rendering mechanisms are similar in these app stores, our experimental results would also work.

\textbf{Internal Validity:} 
Firstly, the ad-related mobile costs we have studied are derived from user reviews. We consider that user reviews can be utilized to identify app issues. Although the cost aspects may not be complete, we obtain a lower bound for the ad-related issues. Secondly, in the paper, we have only analyzed 12 ad integration schemes, whereas app developers would embed ads provided by other ad networks or display different ad formats. We focus on studying the performance costs of different schemes in the paper, and attempt to provide app developers suggestions on better incorporating ads. Developers can use our framework $\mathsf{IntelliAd}$ to detect the costs when using different ad networks or ad formats. Thirdly, not all the ad formats (\textit{e.g.}, video ads) are examined in the paper. This is because we have not inspected any activity invoking the video ad display. We do consider the video-ad activity during the static analysis. Finally, the rendering duration of ads may affect the costs measured. We employ dynamic analysis to monitor the whole process and thereby each ad has the same displaying period. Furthermore, to ensure the reliability of our results, we have repeated the measurements four times. There are some other factors (\textit{e.g.}, ad placement) which may impact the user perceptions, but we suppose that the factor has little influence on the performance costs examined (\textit{i.e.}, memory/CPU overhead, traffic usage, and battery consumption). We leave this as future work for providing better user experience.

\section{Related Work}\label{sec:literature}
In this section, we present two lines of work that inspire our work on ad analysis, namely ad cost identification, and ad cost improvement. 

\subsection{Ad Cost Identification} \label{adiden}
Mobile ads can generate several types of costs for end users, \textit{e.g.}, battery power drainage, memory overhead, CPU utilization, traffic consumption, and privacy leakage, etc. By monitoring and profiling the network usage and system calls related to mobile ads, Wei \textit{et al.}~\cite{wei2012profiledroid} discover that the ``free'' nature in free apps comes with a hard-to-quantify, but noticeable, user cost, such as the profit-making extra traffic (\textit{e.g.}, ads and analytics). Sandra \textit{et al.}~\cite{soroa2010factors} employ a questionnaire survey method to study what factors are most effective in predicting the user response to mobile advertising, and discover that misperceptions of social norms can influence such response. Prashanth \textit{et al.}~\cite{mohan2013prefetching} focus on prefetching mobile ads to alleviate the energy overhead. Similarly, Narseo \textit{et al.}~\cite{vallina2012breaking} characterize mobile ad traffic, and develop a system for enabling energy efficient and network friendly cache-based ad delivery. Suman~\cite{nath2015madscope} investigates what targeting information mobile apps send to ad networks, and how effectively ad networks utilize the information for targeting users. In~\cite{ruiz2014relationship}, Israel \textit{et al.} conclude that the number of ad libraries in an app is not associated with the app rating. Different from their work, we concentrate on the impact of ad integration schemes on the ad costs.

In~\cite{kong2015autoreb}, user reviews have been utilized to learn the security/privacy related behavior, such as spamming, financial issue, over-privileged permission, and data leakage. The work verifies that user reviews can help predict the security issues. Similarly, we regard user reviews as expressions of user perceptions, but we focus on studying the issues related to ad costs. Furthermore, we implement an automatic framework to measure the costs, and verify the measured results with the user feedback in the end. As~\cite{khan2012mitigating} states, the current ``advertisement-supported mobile application model'' is unsustainable due to the tensions between the user preferences for ``free content'', the app developer's desire for revenue, and the ad network's need to display ads. We aim at providing app developers suggestions on better integrating ads, and meanwhile, preventing the shrink of user volume caused by ads.

\subsection{Ad Cost Improvement} \label{adimp}
An amount of research strives to protect users' benefits. Na \textit{et al.}~\cite{wang2015investigating} investigate users' behaviors when the ads library control services are provided. DECAF~\cite{liu2014decaf}, AFrame~\cite{zhang2013aframe}, AdSplit~\cite{shekhar2012adsplit}, and PUMA~\cite{hao2014puma} can isolate ads from the app content based on ad display or code execution, and therefore detect ad fraud, and reduce privacy leakage. Jiaping \textit{et al.}~\cite{gui2015truth} conduct an experimental study on the hidden costs caused by ad rendering, including network usage, app runtime performance, and battery power. They emphasize that the ostensibly free apps can bring users hidden costs, and inform app developers to weigh the tradeoffs of incorporating ads into the mobile apps. Different from their work, ours is established on the premise that ad rendering is integral to the app development. We are not aiming at proving that ad incorporating leads to the hidden costs. We are interested in how to better embed ads and ensure the user experience at the same time.

\section{Conclusions}\label{sec:conclusion}
This paper aims at discovering the ad costs concerned by users, and providing the developers suggestions on integrating ads into apps while ensuring the user experience. We discover four types of ad costs concerning users, \textit{i.e.}, number of ads, memory/CPU overhead, traffic usage, and battery consumption. Different from the previous studies which are app-based, we examine the ad costs based on the ad integration schemes. We capture 12 ad integration schemes of 104 popular Android apps, and then design a framework named $\mathsf{IntilliAd}$ to automatically measure the ad costs of different ad schemes. We demonstrate that different ad schemes indeed generate different costs. Then we explore the correlations between our measured costs and user perceptions. We discover that there exist both correlations and gaps between the measurement results and user feelings. Users are more sensitive to the costs such as memory/CPU overhead, and express little attention about the traffic usage.




{
\bibliographystyle{abbrv}
\bibliography{sigproc}  

\begin{thebibliography}{10}

\bibitem{adreport}
{A Hand-Held World: The Future of Mobile Advertising}.
\newblock
  \url{http://www.business.com/mobile-marketing/the-future-of-mobile-advertising/}.

\bibitem{adlibrary}
{Ad libraries provided by AppBrain.}
\newblock \url{http://www.appbrain.com/stats/libraries/ad}.

\bibitem{admob}
{AdMob}.
\newblock \url{https://www.google.com/admob/}.

\bibitem{androiddrawer}
{AndroidDrawer.}
\newblock \url{http://www.androiddrawer.com/}.

\bibitem{apktool}
{Apktool.}
\newblock \url{http://ibotpeaches.github.io/Apktool/}.

\bibitem{atandt}
{AT\&T}.
\newblock \url{https://www.att.com/}.

\bibitem{onepageads}
{Behavioural policies.}
\newblock \url{https://support.google.com/admob/answer/2753860?hl=en-GB}.

\bibitem{adformat}
{Choose the right ad format.}
\newblock \url{https://www.google.com/admob/monetize.html/}.

\bibitem{cputime}
{CPU time.}
\newblock \url{https://en.wikipedia.org/wiki/CPU_time}.

\bibitem{addisplay}
{Dispay ads}.
\newblock \url{https://www.comscore.com/Insights/Press-Releases/2012/4/}.

\bibitem{goodnetwork}
{How to choose a good mobile ad network.}
\newblock \url{http://www.avocarrot.com/blog/choose-good-mobile-ad-network/}.

\bibitem{inmobi}
{InMobi}.
\newblock \url{http://china.inmobi.com/}.

\bibitem{calculator}
{Mobile advertising calculator.}
\newblock \url{http://ryanmorel.com/mobile-advertising-calculator/}.

\bibitem{mopub}
{MoPub}.
\newblock \url{http://www.mopub.com/}.

\bibitem{paid}
{PAID}.
\newblock \url{http://appsrv.cse.cuhk.edu.hk/~cygao/works/paid.html}.

\bibitem{tapjoy}
{Rewarded Advertising Boosts In-App Ad Sentiment and Brand Perceptions, New
  Study Finds}.
\newblock
  \url{https://home.tapjoy.com/newsroom/press-releases/rewarded-advertising-boosts-inapp-ad-sentiment-and-bra/}.

\bibitem{inappad}
{Think the app vs. mobile web battle is over?}
\newblock \url{http://venturebeat.com/2015/07/28/}.

\bibitem{whitepaper}
{Understanding today’s smartphone user}.
\newblock
  \url{http://www.informatandm.com/wp-content/uploads/2012/02/Mobidia_final.pdf}.

\bibitem{ram}
{Why does the iPhone need so much less RAM than Android devices?}
\newblock \url{https://www.quora.com/}.

\bibitem{chen2014ar}
N.~Chen, J.~Lin, S.~C. Hoi, X.~Xiao, and B.~Zhang.
\newblock Ar-miner: mining informative reviews for developers from mobile app
  marketplace.
\newblock In {\em Proceedings of the 36th International Conference on Software
  Engineering (ICSE)}, pages 767--778. ACM, 2014.

\bibitem{paid2015cygao}
C.~Gao, B.~Wang, P.~He, J.~Zhu, Y.~Zhou, and M.~R. Lyu.
\newblock Paid: Prioritizing app issues for developers by tracking user reviews
  over versions.
\newblock In {\em Proceedings of the 26th International Symposium on Software
  Reliability Engineering (ISSRE)}. IEEE, 2015.

\bibitem{gu2015parts}
X.~Gu and S.~Kim.
\newblock " what parts of your apps are loved by users?"(t).
\newblock In {\em Proceedings of the 30th International Conference on Automated
  Software Engineering (ASE)}, pages 760--770. IEEE, 2015.

\bibitem{gui2015truth}
J.~Gui, S.~Mcilroy, M.~Nagappan, and W.~G. Halfond.
\newblock Truth in advertising: The hidden cost of mobile ads for software
  developers.
\newblock In {\em Proceedings of the 37th International Conference on Software
  Engineering (ICSE)}, pages 100--110. IEEE, 2015.

\bibitem{hao2014puma}
S.~Hao, B.~Liu, S.~Nath, W.~G. Halfond, and R.~Govindan.
\newblock Puma: Programmable ui-automation for large-scale dynamic analysis of
  mobile apps.
\newblock In {\em Proceedings of the 12th International Conference on Mobile
  Systems, Applications, and Services (MobiSys)}, pages 204--217. ACM, 2014.

\bibitem{khalid2014mobile}
H.~Khalid, E.~Shihab, M.~Nagappan, and A.~Hassan.
\newblock What do mobile app users complain about? a study on free ios apps.
\newblock 2014.

\bibitem{khalid2015mobile}
H.~Khalid, E.~Shihab, M.~Nagappan, and A.~E. Hassan.
\newblock What do mobile app users complain about?
\newblock {\em Software, IEEE}, 32(3):70--77, 2015.

\bibitem{khan2012mitigating}
A.~J. Khan, V.~Subbaraju, A.~Misra, and S.~Seshan.
\newblock Mitigating the true cost of advertisement-supported free mobile
  applications.
\newblock In {\em Proceedings of the 12th Workshop on Mobile Computing Systems
  \& Applications (HotMobile)}, page~1. ACM, 2012.

\bibitem{kong2015autoreb}
D.~Kong, L.~Cen, and H.~Jin.
\newblock Autoreb: Automatically understanding the review-to-behavior fidelity
  in android applications.
\newblock In {\em Proceedings of the 22nd SIGSAC Conference on Computer and
  Communications Security (CCS)}, pages 530--541. ACM, 2015.

\bibitem{liu2014decaf}
B.~Liu, S.~Nath, R.~Govindan, and J.~Liu.
\newblock Decaf: detecting and characterizing ad fraud in mobile apps.
\newblock In {\em Proceeding of the 11th USENIX Symposium on Networked Systems
  Design and Implementation (NSDI)}, pages 57--70, 2014.

\bibitem{mohan2013prefetching}
P.~Mohan, S.~Nath, and O.~Riva.
\newblock Prefetching mobile ads: Can advertising systems afford it?
\newblock In {\em Proceedings of the 8th European Conference on Computer
  Systems}, pages 267--280. ACM, 2013.

\bibitem{nath2015madscope}
S.~Nath.
\newblock Madscope: Characterizing mobile in-app targeted ads.
\newblock In {\em Proceedings of the 13th International Conference on Mobile
  Systems, Applications, and Services (MobiSys)}, pages 59--73. ACM, 2015.

\bibitem{rastogi2013appsplayground}
V.~Rastogi, Y.~Chen, and W.~Enck.
\newblock Appsplayground: automatic security analysis of smartphone
  applications.
\newblock In {\em Proceedings of the 3rd Conference on Data and Application
  Security and Privacy (CODASPY)}, pages 209--220. ACM, 2013.

\bibitem{ruiz2014relationship}
I.~M. Ruiz, M.~Nagappan, B.~Adams, T.~Berger, S.~Dienst, and A.~Hassan.
\newblock On the relationship between the number of ad libraries in an android
  app and its rating.
\newblock {\em IEEE Software}, 99(1), 2014.

\bibitem{shekhar2012adsplit}
S.~Shekhar, M.~Dietz, and D.~S. Wallach.
\newblock Adsplit: Separating smartphone advertising from applications.
\newblock In {\em Proceedings of the 21st USENIX Security Symposium (USENIX
  Security)}, pages 553--567, 2012.

\bibitem{soroa2010factors}
S.~Soroa-Koury and K.~C. Yang.
\newblock Factors affecting consumers’ responses to mobile advertising from a
  aocial norm theoretical perspective.
\newblock {\em Telematics and Informatics}, 27(1):103--113, 2010.

\bibitem{vallina2012breaking}
N.~Vallina-Rodriguez, J.~Shah, A.~Finamore, Y.~Grunenberger, K.~Papagiannaki,
  H.~Haddadi, and J.~Crowcroft.
\newblock Breaking for commercials: characterizing mobile advertising.
\newblock In {\em Proceedings of Conference on Internet Measurement
  Conference}, pages 343--356. ACM, 2012.

\bibitem{vu2015mining}
P.~M. Vu, T.~T. Nguyen, H.~V. Pham, and T.~T. Nguyen.
\newblock Mining user opinions in mobile app reviews: A keyword-based approach
  (t).
\newblock In {\em Proceedings of the 30th International Conference on Automated
  Software Engineering (ASE)}, pages 749--759. IEEE, 2015.

\bibitem{wang2015investigating}
N.~Wang, B.~Zhang, B.~Liu, and H.~Jin.
\newblock Investigating effects of control and ads awareness on android users'
  privacy behaviors and perceptions.
\newblock In {\em Proceedings of the 17th International Conference on
  Human-Computer Interaction with Mobile Devices and Services (MobileHCI)},
  pages 373--382. ACM, 2015.

\bibitem{wei2012profiledroid}
X.~Wei, L.~Gomez, I.~Neamtiu, and M.~Faloutsos.
\newblock Profiledroid: multi-layer profiling of android applications.
\newblock In {\em Proceedings of the 18th International Conference on Mobile
  Computing and Networking (MobiCom)}, pages 137--148. ACM, 2012.

\bibitem{yoon2012appscope}
C.~Yoon, D.~Kim, W.~Jung, C.~Kang, and H.~Cha.
\newblock Appscope: Application energy metering framework for android
  smartphone using kernel activity monitoring.
\newblock In {\em Presented as part of the 2012 USENIX Annual Technical
  Conference (USENIX ATC)}, pages 387--400, 2012.

\bibitem{zhang2013aframe}
X.~Zhang, A.~Ahlawat, and W.~Du.
\newblock Aframe: Isolating advertisements from mobile applications in android.
\newblock In {\em Proceedings of the 29th Annual Computer Security Applications
  Conference (ACSAC)}, pages 9--18. ACM, 2013.

\end{thebibliography}
\par
} 

\end{document}